\documentclass[%
aip,
sd,%
amsmath,amssymb,
preprint,%
]{revtex4-1}

\usepackage{graphicx}% Include figure files
\usepackage{dcolumn}% Align table columns on decimal point
\usepackage{bm}% bold math
\usepackage{wasysym}
\usepackage{setspace}
\usepackage{color}
\usepackage{nicefrac}
\usepackage{times}

\newcommand{\figref}[1]{Fig.~\ref{fig:#1}}
\newcommand{\Figref}[1]{Figure~\ref{fig:#1}}

\renewcommand{\eqref}[1]{(\ref{eq:#1})}

\newcommand{\eqreftwo}[2]{(\ref{eq:#1}) and (\ref{eq:#2})}

\newcommand{\Eqref}[1]{Equation~(\ref{eq:#1})}
\newcommand{\citeasnoun}[1]{Ref.~\onlinecite{#1}}

\newcommand{\Td}{T_\text{d}}
\newcommand{\Te}{T_\text{e}}
\newcommand{\kB}{k_{\mathrm{B}}}

\newcommand{\da}{\delta a}

\renewcommand{\ge}{\gamma_\text{e}}
\newcommand{\gd}{\gamma_\text{d}}
\newcommand{\xe}{\xi_\text{e}}
\newcommand{\xd}{\xi_\text{d}}
\renewcommand{\sp}{s_\text{p}}
\renewcommand{\wp}{\omega_\text{p}}
\newcommand{\zetap}{\zeta}
\newcommand{\ds}{|\delta s_{-}|^2}

\begin{document}

\preprint{AIP/123-QED}

\title{Thermal radiation from optically driven Kerr ($\chi^{(3)}$)
  photonic cavities}

\author{Chinmay Khandekar}
\affiliation{Department of Electrical Engineering, Princeton University, Princeton, NJ 08540}
\author{Zin Lin}
\affiliation{School of Engineering and Applied Sciences, Harvard University, Cambridge, MA 02139}
\author{Alejandro W. Rodriguez}
\affiliation{Department of Electrical Engineering, Princeton University, Princeton, NJ 08540}

\begin{abstract}
  We study thermal radiation from nonlinear ($\chi^{(3)}$) photonic
  cavities coupled to external channels and subject to incident
  monochromatic light. Our work extends related work on nonlinear
  mechanical oscillators [Phys. Rev. Lett. 97, 110602 (2006)] to the
  problem of thermal radiation, demonstrating that bistability can
  enhance thermal radiation by orders of magnitude and result in
  strong lineshape alternations, including ``super-narrow spectral
  peaks'' ocurring at the onset of kinetic phase transitions.  We show
  that when the cavities are designed so as to have perfect linear
  absorptivity (rate matching), such thermally activated transitions
  can be exploited to dramatically tune the output power and radiative
  properties of the cavity, leading to a kind of Kerr-mediated
  thermo-optic effect. Finally, we demonstrate that in certain
  parameter regimes, the output radiation exhibits Stokes and
  anti-Stokes side peaks whose relative magnitudes can be altered by
  tuning the internal temperature of the cavity relative to its
  surroundings, a consequence of strong correlations and interference
  between the emitted and reflected radiation.
\end{abstract}

\pacs{}

\keywords{}

\maketitle

Driven nonlinear oscillators, including optical,~\cite{Vahala03}
optomechanical,~\cite{Kippenberg:07} and MEMS~\cite{Cross08,Quidant13}
resonators, have been studied for decades and exploited for many
applications, from mass detection~\cite{Chaste12} to
sensing~\cite{Cleland02} and tunable filtering.~\cite{Buks06} When
driven to a non-equilibrium state, these systems can exhibit a wide
range of unusual thermal phenomena,~\cite{Moser12} leading for
instance to cooling and amplification of thermal fluctuations in
optomechanical systems,~\cite{Kippenberg:07} generation of squeezed
states of light in Kerr media,~\cite{Wu87} and stochastic
resonances.~\cite{Marchesonl98} Previous studies of Duffing
oscillators have also identified novel effects arising from the
nonlinear interaction of coherent pumps with thermal
noise,~\cite{dykman75a,dykman92,dykman94} leading to phase transitions
and lineshape alterations that were recently observed in a handful of
systems, e.g.  mechanical oscillators~\cite{HoChan06,Stambaugh06} and
Josephson junctions.~\cite{Andre12}

In this letter, we study thermal radiation effects in optically driven
$\chi^{(3)}$ photonic cavities coupled to external channels.  We
demonstrate that in certain parameter regimes,
bistability~\cite{Notomi05,Cowan03} in photonic resonators leads to
thermally activated transitions that amplify thermal fluctuations by
orders of magnitude and cause dramatic changes in the cavity spectrum,
analogous to noise-induced switching in mechanical
oscillators.~\cite{Stambaugh06} We find that when the photonic cavity
is critically coupled to the radiation channel (enforced by designing
the cavity to have equal dissipation and radiation
rates),~\cite{JoannopoulosJo08-book} the coherent part of the output
power varies dramatically with temperature, leading to a kind of
Kerr-mediated thermo-optic effect. A simple perturbative analysis also
shows that outside of the bistability region, the interaction of the
coherent drive with thermal noise leads to amplified, Raman-type
Stokes and anti-Stokes side peaks in the radiation spectrum, the
relative amplitudes of which depend on a sensitive interference
between the externally incident and reflected thermal
radiation. Related phenomena have been long studied~\cite{dykman94}
and more recently observed~\cite{HoChan06,Stambaugh06} in the context
of driven nonlinear mechanical oscillators as well as resonators based
on rf-driven Josephson junctions~\cite{Andre12}, and microscopic
theories have also been used to describe related optical effects in
the quantum regime.~\cite{Walls80} Our work is an extension of these
studies to the particular problem of thermal radiation from photonic
resonators. As we show below, additional considerations arising in the
case of radiation from cavities but absent in mechanical oscillators
or bulk media, such as the strong coupling of the cavity to an
external channel, dramatically impact the outgoing radiation. The
ability to tune the radiation properties of resonators via temperature
and optical signals offers potentially new avenues for applications in
nano-scale heat regulation,~\cite{WangLi08} detection,~\cite{Notomi13}
rectification,~\cite{Otey10,Roberts11} photovoltaics,~\cite{lenert14}
or incoherent sources.~\cite{Noda07} We propose a practical, photonic
structure where these effects can arise near room temperature and at
mW powers.

% TODO: like thermo-optic effect, here we show that temperature can
% modulate optical properties but this time optical nonlinearities
% rather than phonons responsible, and we show power tunability on the
% order of $\eta^2_T / \gamma^2$, which for realistic
% materials/cavities corresponds to 0.01~mW/K.

\begin{figure}[t!]
\centering
\includegraphics[width=0.55\linewidth]{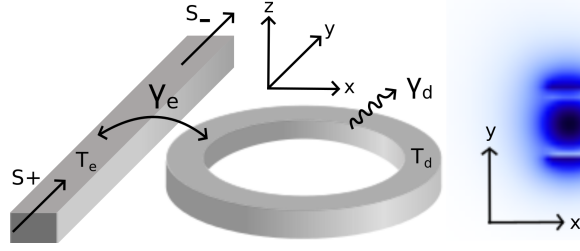}
\caption{Schematic of a wavelength-scale silicon ring resonator of
  radius $R=4.4\mu$m, height $h=220$nm, and width $w=350$nm, coupled
  to a silicon waveguide (channel), both on a silica substrate. Also
  shown is the $E_y$ mode profile of a resonance designed to have
  azithmutal number $m=25$, wavelength $\lambda=1.5\mu$m, radiative
  lifetimes $\gtrsim 10^6$, and relatively large nonlinear coupling
  coefficient $\alpha = 0.032 \chi^{(3)}\omega_0/(8 \epsilon_0
  \lambda^3)$. The loss $\gd$ and waveguide--coupling $\ge$ rates are
  much larger than the corresponding radiation rate.}
\label{fig:design}
\end{figure}

The system under consideration belongs to the class of nonlinear
photonic resonators depicted in \figref{design}, involving a cavity
coupled to an external channel (e.g. a waveguide). The description of
thermal radiation in this system can be carried out via the
coupled-mode theory framework~\cite{Haus84:coupled}, which we recently
employed to study thermal radiation in a related
system~\cite{chinmay15} but now extend to consider the addition of a
coherent pump. The equations describing the cavity mode $a$ are given
by:~\cite{chinmay15}
\begin{eqnarray}
\label{eq:cavity1}
\frac{da}{dt} &=& [i(\omega_{0} - \alpha|a|^{2})-\gamma]a + \sqrt{2\gd} \xd + \sqrt{2\ge} s_{+}, \\
s_{-} &=& -s_{+} + \sqrt{2\ge} a,
\label{eq:cavity2}
\end{eqnarray}
where $|a|^{2}$ is the energy of the cavity mode and $|s_\pm|^2$ are
the incident ($+$) and outgoing ($-$) power from and into the external
channel, respectively. The latter arises due to dissipative noise
inside the cavity $\xd$ as well as externally incident light $s_+$
consisting of both thermal radiation $\xe$ and a monochromatic pump
$\sp \exp(i\wp t)$. The dynamics of the cavity field are described by
its resonance frequency $\omega_0$ and decay rate $\gamma=\ge+\gd$,
which includes linear absorption $\gd$ as well as decay into the
external channel $\ge$. The real and imaginary parts of the nonlinear
coefficient $\alpha = \frac{3}{4}\omega_0 \int \varepsilon_0
\chi^{(3)} |\vec{E}|^4 / (\int \varepsilon |\vec{E}|^2)^2$ depend on a
complicated overlap integral of the linear cavity
fields,~\cite{Rodriguez07:OE} and lead to self-phase modulation (SPM)
and two-photon absorption (TPA), respectively. We mainly focus on the
effects of SPM (real $\alpha > 0$) since we find that TPA leads to
thermal broadening of the kind explored in~\citeasnoun{chinmay15}.
Both internal and external thermal sources are represented by
stochastic, delta-correlated white-noise sources $\xe$ and $\xd$
satisfying (assuming $\gamma \ll \omega_0$),
\begin{align}
\label{eq:sfdt}
\langle \xi^{*}(t) \xi(t') \rangle = \Theta(\omega_0,T)
\delta(t-t'),
\end{align}
where $\langle \ldots \rangle$ denotes a thermodynamic or ensemble
average, and $\Theta(\omega,T)=\hbar\omega / (e^{\hbar\omega/\kB
  T}-1)$ is the mean energy of a Planck oscillator~\cite{Otey10} at
local temperature $T$; the temperatures of the internal and external
baths are denoted as $\Td$ and $\Te$, respectively. Above, we assumed
$\hbar \gamma/\kB \ll \Te,\Td$ allowing us to ignore the
frequency-dispersion and temporal correlations (colored noise)
associated with $\Theta$. (Note that in the limit $\hbar \omega_0 /
k_\text{B} T \to 0$ one obtains the classical result $\Theta \to
k_\text{B} T$.)

% To begin with, we review features related to absorption in bistable
% photonic cavities which, when combined with thermal fluctuations,
% lead fluctuations. To begin with, we write down the steady-state
% cavity energy obtain the zeroth order solution $a_0$ due to the
% coherent drive

% NOTE: as expected, even though cavity exhibits hysterisis when
% coherently pumped, the system is completely reversible with respect
% to $T$.

\begin{figure*}[t!]
\centering
\includegraphics[width=0.95\linewidth]{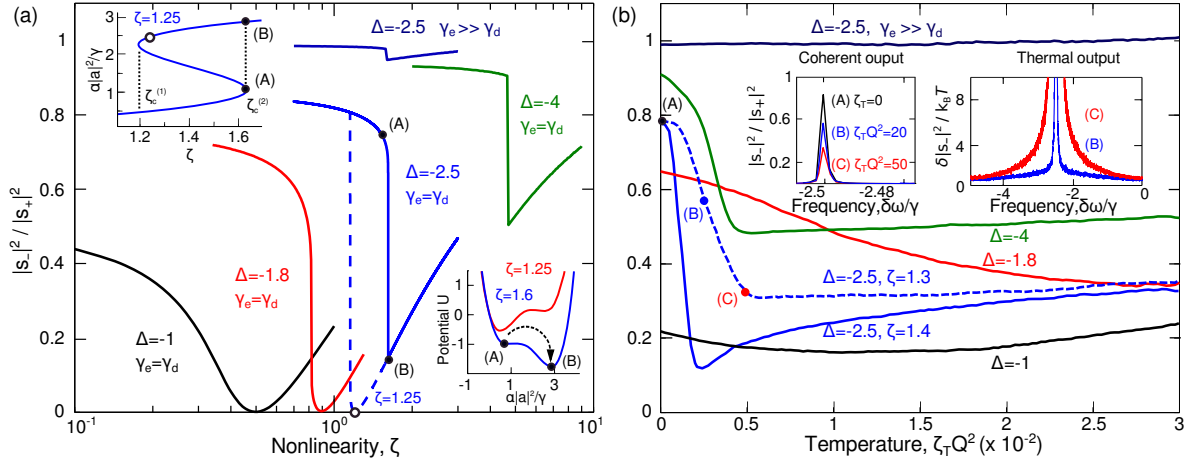}
\caption{(a) Output power $|s_{-}|^2$ normalized by the input power
  $|s_{+}|^2$ of the pumped system described in \figref{design}, in
  the absence of thermal noise and as a function of $\zetap$, for
  different values of detuning
  $\Delta=\frac{\wp-\omega_0}{\gamma}$. The top left inset shows a
  hysterisis plot of the energy $\alpha |a_0|^2/\gamma$ as a function
  of $\zetap$, the solutions of \eqref{cubic}, for the particular
  choice of $\Delta=-2.5$ while the bottom inset shows the
  corresponding potential energy $U$ as a function of the cavity
  energy for two different $\zetap=|\Delta/2|$ and $\zetap^{(2)}$. (b)
  The same normalized output power $|s_{-}|^2/|s_{+}|^2$ as a function
  of temperature $\zeta_T Q^2$, where $\zeta_T = \alpha
  \Theta(\omega_0,T) \ge/\gamma^2$ and $Q=\omega_0/\gamma$, for
  different values of $\Delta$ and $\zeta \lesssim \zeta^{(2)}$. The
  insets illustrate the change in the coherent (left) and thermal
  radiation (right) spectra. Both internal and external baths have
  equal temperatures $\Td=\Te=T$.}
\label{fig:transition}
\end{figure*}

\emph{Thermal amplification and power tunability.---} We show that
bistability can amplify thermal fluctuations and lead to enhanced,
temperature--tunable emission from the cavity. We begin by reviewing a
number of key features of the system in the absence of thermal noise,
whose contributions are considered perturbatively due to the generally
weak nature of thermal noise, $|\sp|^2 \gg \gamma \kB T$. The
steady-state cavity field $a_0$ due to the pump is given by the
well-known cubic equation:~\cite{Marin02,dykman94}
\begin{align}                                                                 
  \left[\left(\Delta + \frac{\alpha|a_0|^2}{\gamma}\right)^2 +
    1\right]\frac{\alpha|a_0|^2}{\gamma} = 2\zetap,
  \label{eq:cubic}
\end{align}
where $\zetap \equiv \alpha |\sp|^2 \ge/ \gamma^3$ is the effective
nonlinear coupling associated with the pump and $\Delta \equiv
\frac{\wp-\omega_0}{\gamma}$ is the dimensionless
detuning. \Eqref{cubic} describes a number of extensively studied
nonlinear effects,~\cite{Marin02,Johnson01:cavities} including
bistability arising in the regime $\Delta < -\sqrt{3}$ and
$\zetap^{(1)} < \zetap < \zetap^{(2)}$, as illustrated by the
hysterisis plot on the inset of \figref{transition}(a) which shows the
dimensionless cavity energy $x=\alpha |a|^2/\gamma$ as a function of
$\zeta$.

An effect that seems little explored but that plays an important role
on the thermal properties of this system is perfect absorption, which
occurs when a photonic cavity is driven on resonance and its
dissipation and radiation rates are equal, also known as rate
matching.~\cite{JoannopoulosJo08-book} In the presence of
nonlinearities, the cavity frequency and hence the absorbed power
depend on $\zetap$. For instance, in the non-bistable regime, the
output power varies slowly with $\zetap$, as illustrated by the green
curve in \figref{transition}(a) for $\Delta =-1$, increasing and then
decreasing as $\zetap \to |\Delta/2|$, at which point the cavity and
pump frequencies are in resonance, i.e. $\alpha |a_0|^2/\gamma =
-\Delta$.  Bistability can lead to a more pronounced dependence on
$\zetap$: the two stable steady states experience different frequency
shifts and hence loss rates, and ultimately which state is excited in
the steady state depends on the specific initial (or excitation)
conditions.~\cite{Strogatz94} \Figref{transition}(a) shows the
steady-state output power $|s_{-}|^2$ as $\zetap$ is adiabatically
increased (solid lines) from zero and above the critical point
$\zetap^{(2)}$, for multiple $\Delta$. The dashed blue line shows the
power as $\zetap$ is adiabatically decreased below $\zetap^{(2)}$ for
the particular case $\Delta=-2.5$, demonstrating that only the upper
branch experiences perfect absorption, occuring at $\zetap=|\Delta/2|$
and marked by the white circle. The corresponding change in the output
power as the system transitions from the lower (A) to the higher (B)
energy state at $\zetap^{(2)}$ is given approximately by:
\begin{align}
%  \frac{|s_{-}|^2_B - |s_{-}|^2_A}{|s_{+}|^2} \approx
 |s_{+}|^2 \bigg(1-\frac{\ge-\gd}{\ge+\gd}\bigg) \frac{(\Delta+x_1)^2 -
    (\Delta+x_2)^2}{[1+(\Delta+x_1)^2][1+(\Delta+x_2)^2]}
  \label{eq:jump}
\end{align}
where $x_1=-\frac{1}{3}(2\Delta+\sqrt{\Delta^2-3})$ and $x_2=-2
(\Delta+x_1)$ are the cavity energies associated with the lower and
higher energy state, respectively. Given \eqref{jump}, one can show
that the difference in output power is largest under the rate matching
condition $\ge=\gd$ and at $\Delta \approx -7/3$, decreasing with
smaller or larger detuning.

The presence of noise complicates this picture due to
finite--temperature fluctuations which cause the system to undergo
transitions between the two states, where the rates of
forward/backward transitions are a complicated function of the
potential energy $U$ and temperature of the
system.~\cite{dykman94,Strogatz94} (For convenience and without loss
of generality, we take both thermal baths to have the same temperature
$T$.)  In particular, $U$ is obtained by integrating the steady-state
equation associated with the cavity energy $\frac{dx}{dt} = [(\Delta +
x)^2 + 1] x - 2\zeta = 0$ with respect to $x=\alpha
|a|^2/\gamma$. Examples of $U$ are shown on the lower inset of
\figref{transition}(a) for two values of $\zeta$. Thermally activated
hopping leads to significant enhancement of amplitude fluctuations,
which manifest as large changes in the radiation spectrum of the
cavity. This is illustrated by the top inset of
\figref{transition}(b), which shows the thermal spectrum of the power
$|\delta s_{-}(\omega)|^2$ for the particular choice of $\Delta=-2.5$
and $\zetap=1.3$ and for multiple values of $\zeta_T Q^2$, where for
convenience (below) we have introduced the dimensionless effective
thermal coupling $\zeta_T\equiv \alpha \Theta(\omega_0,
T)\ge/\gamma^2$ and cavity-lifetime
$Q=\omega_0/\gamma$.~\cite{JoannopoulosJo08-book} Such enhancements
were predicted to occur and recently observed in nonlinear mechanical
oscillators,~\cite{dykman94,HoChan06} where the authors showed that at
special $\zetap$, the system undergoes a so-called kinetic phase
transition associated with equal rates of forward/backward hopping and
exhibits a ``supernarrow'' and highly amplified spectral
peak. Interestingly, we find that in the case of optical resonators,
thermal amplification can be accompanied by a significant decrease in
the coherent output power despite the fact that $|\sp|^2 \gg \gamma
\kB T$, a consequence of perfect absorption. In particular, operating
under rate matching and near $\zetap^{(2)}$ allows for temperature to
initiate transitions $(\text{A})\rightleftharpoons(\text{B})$, leading
to significant changes in $|s_{-}|^2$ with respect to
$T$. Essentially, as $\zetap \to \zetap^{(2)}$, the potential barrier
separating the lower $x_1$ from the higher $x_2$ energy states begins
to dissapear, resulting in increased rate of forward transitions and
hence larger absorption.

These features are illustrated in \figref{transition}(b) which shows
the total output power as a function of $\zeta_T Q^2$ for different
combinations of $\zeta$ and $\Delta$. (We found numerically that for a
given $\zeta$ and $\Delta$, changing either $\zeta_T$ or $Q^2$ while
leaving $\zeta_T Q^2$ unchanged leaves $|s_{-}|^2/|s_{+}|^2$
unaltered.)  Noticeably, while the change in the output power is
gradual in the non-bistable regime ($\Delta > -\sqrt{3}$), there is a
significantly stronger dependence in the bistable regime---the slope
becomes increasingly sharper as $\zetap \to \zetap^{(2)}$ and $\zeta_T
\to 0$ since it becomes increasingly easier for lower $T$ fluctuations
to induce hopping unto the higher-energy state. At sufficiently large
$\zeta_T$, $|s_{-}|^2$ is found to increase with increasing $\zeta_T$
as the cavity field no longer probes the hysterisis regime.  While the
maximum change in $|s_{-}|^2$ can be estimated from the steady-state
analysis in the absence of noise (with the largest change occuring for
$\Delta \approx -7/3$), its dependence on $\zeta_T$ is a complicated
function of $\zeta$ and $\Delta$. For instance, for $\Delta=-2.5$ and
$\zetap=1.4$ [blue line in \figref{transition}(b)], the sharp decrease
in output power occurs at $\zeta_T Q^2 \approx 10$ and yields a slope
$\frac{1}{|s_{+}|^2} \frac{\delta(|s_-|^2)}{\delta(\zetap_T Q^2)}
\approx 0.05$.

\emph{Side peaks.---} We now show that the radiation spectrum also
exhibits other interesting features, including the emergence of
Raman-type Stokes and anti-Stokes side peaks previously observed in
driven mechanical oscillators.~\cite{HoChan06,dykman94} Interestingly,
we find that in our photonic resonator, the presence of the external
channel dramatically alters the relative amplitudes of the side peaks,
e.g. leading to a symmetric spectrum when the two baths have equal
temperatures. We begin by exploiting a simple perturbation theory in
which the thermal fluctuations of the cavity-field $\da$ and radiation
$\delta s_{-} = -\xe + \sqrt{2\ge} \da$ are treated perturbatively,
leading to analytical expressions for the corresponding thermal energy
and radiation spectra. Assuming $|\sp|^2 \gg \gamma \kB \{\Td,\Te\}$,
we obtain:
\begin{align}
  \label{eq:PSDthermal} 
  \langle |\da|^2 \rangle &= \frac{\kB f_+ (2\gd \Td + 2\ge T_e)}{\
    D} \\
  \langle |\delta s_{-}|^2 \rangle &= \kB \Te + \frac{4\ge \kB}{D} [
  \gd (f_{+}T_{d}-f_{-}T_{e}) +\ge(f_{+}-f_{-})T_{e}]
\label{eq:PSDoutput}
\end{align}
where $f_{+},f_{-}$ and $D$ are given by
\begin{align*} 
  f_{\pm} &= (\omega + \omega_0 - 2 \wp - 2\alpha |a_0|^2)^2 +
  \gamma^2 \pm \
  \alpha^2|a_0|^4 \\
  D &= [\gamma^2 + (\omega_0 - \wp - 2\alpha|a_0|^2)^2 - (\omega -
  \wp)^2 - \alpha^2|a_0|^4]^2 \\ &+ 4\gamma^2(\omega-\wp)^2,
\end{align*}
and where $|a_0|^2$ denotes the steady-state cavity energy in the
absence of fluctuations, the solution of \eqref{cubic}. Here, for
simplicity we have assumed the classical limit $\Theta(\omega_0,T)
\rightarrow \kB T$. In the absence of the external bath $T_e=0$, the
above equations are similar to those obtained in the case of
mechanical oscillators.~\cite{dykman94} The situation changes with the
channel due to the unavoidable intereference and induced correlations
of the emitted and reflected radiation, described in \eqref{PSDoutput}
by the $f_{-}$ terms.

\begin{figure*}
\centering
\includegraphics[width=0.95\linewidth]{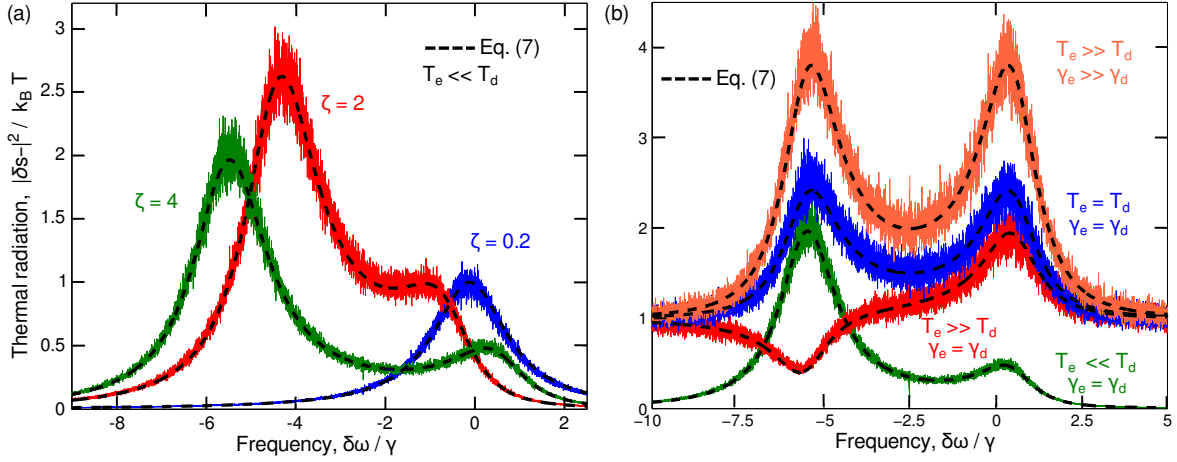}
\caption{Thermal radiation $|\delta s_{-}|^2$ for the system described
  in \figref{design}, normalized by the maximum of the internal and
  external bath temperatures $T=\max\{\Td,\Te\}$, as a function of the
  dimensionless frequency $\delta\omega/\gamma$ where $\delta\omega =
  (\omega-\omega_0)$, with fixed $\Delta=-2.5$ and under various
  operating conditions. The radiation spectrum is shown (a) in the
  limit $\Td\gg\Te$ of negligible externally incident radiation, for
  different values of $\zetap$ and (b) for fixed $\zetap=4$ but
  different $\Td,\Te$ and linear decay rates $\gd,\ge$.}
\label{fig:sidepeak}
\end{figure*}

\Figref{sidepeak} illustrates the radiation spectrum of the cavity
under different operating conditions, showing excellent agreement
between the numerically (noisy) and analytically (dashed lines)
computed spectra. We note that all of the results shown in
\figref{sidepeak} correspond to cavities operating outside of the
bistable regime: although it is possible to obtain a complete and
analytical description of the spectrum based on
\eqreftwo{PSDthermal}{PSDoutput}, such an analysis is difficult in the
bistable regime due to hopping between states, requiring a complicated
description of the transition rates and stationary distributions of
the system.~\cite{dykman94} For instance, in the bistable regime, one
observes the appearance of a temperature-dependent supernarrow
spectral peak whose amplitude decreases with increasing $\zetap$. In
the absence of bistability, a similar but weaker amplification occurs
as $\zeta \to |\Delta/2|$ or equivalently, as the cavity frequency
becomes resonant with the pump. Regardless of regime, at suffiiently
large $\zeta$ (once the cavity resonance has crossed $\wp$),
modulation of the thermal noise by the pump causes the spectrum to
transition from being singly to doubly resonant due to the emergence
of an additional anti-Stokes peak.~\cite{dykman94} In the limit as
$\zeta \to \infty$, both peaks move farther apart and their amplitudes
assymptote to a system-dependent constant.

\Figref{sidepeak}(b) explores the dependence of the peak amplitudes on
various cavity parameters, including $\Td = \Te$, $\Td \gg \Te$, and
$\Td \ll \Te$, corresponding to a resonator that is either at thermal
equilibrium, heated, or cooled with respect to its surroundings,
respectively. When noise entering the system through the external bath
is negligible $\Td \gg \Te$, similar to the previously explored
situation involving mechanical oscillators,~\cite{dykman94} one finds
that the Stokes peak is always much larger than the anti-Stokes peak
(green line). Essentially, for $\alpha > 0$ the cavity nonlinearity
favors down-conversion, as captured by the asymmetric $f_{+}$ terms
above. The peak radiation associated with the Stokes peak can be
readily obtained from \eqref{PSDthermal} in the non-bistable regime
$\Delta \leq -\sqrt{3}$, and is given by $\max \ds =
\frac{4\ge\gd}{\gamma^2} (1 + 2\Delta^2) \kB \Td$, reaching $7\kB\Td$
precisely at the onset of bistability and when $\ge=\gd$.  At larger
$\zeta > |\Delta/2|$, the amplitude of both peaks decreases with
increasing $\zeta$ where, as $\zeta \to \infty$ (not shown), the
amplitude of the Stokes peak $\to \kB \Td$ while the anti-Stokes peak
dissapears. The situation changes dramatically when the noise entering
the system through the external bath cannot be ignored, i.e. $\Te
\gtrsim \Td$. In particular, as observed from \eqref{PSDthermal},
although the cavity spectrum favors Stokes to anti-Stokes conversion
regardless of the relative temperatures or decay rates, we find that
the spectrum of outgoing radiation can be dramatically different
depending on the regime of operation. When $\Te \gg \Td$ where noise
is dominated by external radiation, we find that the anti-Stokes peak
dominates (red line) except when $\ge \gg \gd$, at which point the
spectrum exhibits a symmetric lineshape (orange line). Such a reversal
in relative amplitudes is captured by the $f_{-}$ terms above, which
include correlations and interference between the emitted and
reflected radiation. The maximum radiation in the non-bistable regime
in this case $\max \ds = \kB \Te (1 + \frac{8\ge^2}{\gamma^2} \Delta^2
- \frac{4\ge\gd}{\gamma^2})$, reaching $25\kB\Te$ at the onset of
bistability and when $\ge\gg\gd$ (rather than under rate
matching). Interestingly, we find that when the two baths lie at the
same temperature $\Te=\Td$, both peaks have equal amplitudes
regardless of $\ge/\gd$, though the maximum amplitude in this regime
also occurs in the limit $\ge\gg\gd$. This unexpected symmetrization
of the spectrum arising due to interference effects seems to be a
unique property of thermal radiation in this system. Although previous
work on nonlinear fluctuations in the quantum regime observed similar
peaks in the spectrum, a symmetric spectrum was found to arise only at
zero temperature (a singular point of the theory~\cite{dykman84}) due
to quantum tunneling.~\cite{Walls80}

Although a number of the abovementioned effects have been observed in
mechanical oscillators, they remain unobserved in the context of
thermal radiation where they could potentially be exploited in
numerous applications.~\cite{Noda07,WangLi08,lenert14} As demonstrated
above, the interplay between the internal and externally incident
radiation and the coherent pump leads to new effects in thermal
radiators, including dramatic changes in both the coherent and thermal
output spectrum with temperature, along with temperature--tunable
Stokes and anti-Stokes side peaks. Finally, we conclude by proposing a
realistic, silicon ring resonator design, depicted schematically in
\figref{design}, where one could potentially observe these effects
near room temperature and with operating $Q \sim 10^{5}$ and input
power $|s_p|^2\sim 1$mW, leading to $\alpha|s_p|^2 Q^2 \sim
|\Delta|$. For these parameters we find that $\frac{1}{|s_{+}|^2}
\frac{\delta(|s_-|^2)}{\delta(T)} \sim 0.04\mathrm{K}^{-1}$ at
$T\approx 300$K. Although this is almost two orders of magnitude
smaller in comparison with thermo-optic effects in silicon, which lead
to tunable powers $\sim \mathrm{K}^{-1}$ for the same structure, at
lower temperatures $T\lesssim 100\mathrm{K}$ where the thermo-optic
coefficient is much smaller~\cite{Nawrodt12}, our fluctuation-induced
effects offer significantly better temperature tunability. Other
cavity designs such as the nanobeam cavity described in
Ref.~\onlinecite{Zin14} yield much larger $\alpha$ and allow smaller
$Q$ to be employed, leading to even larger tunability compared to that
obtained via thermo-optic effects.

We are grateful to Mark Dykman for very helpful comments and
suggestions. This work was supported in part by the National Science
Foundation under Grant No. DMR-145483.

%Since TPA should be negligible, we require nonlinear material with
%large FOM such as $Si$ where the resonator can be designed for
%operation at $\lambda = 1.55\mu m$ at $T_{d}=300K$. With the input
%pump drive at $1mW$ and Kerr nonlinearity $\chi^3=2.45\times
%10^{-19}m^2/W.$~\cite{Jalali11}, we obtain $Q \sim 2\times10^{5}$ to
%observe these effects using a simple ring resonator shown
%in~\figref{design}.

\bibliographystyle{unsrt}
\bibliography{photon}

\begin{thebibliography}{10}

\bibitem{Vahala03}
Kerry~J. Vahala.
\newblock Optical microcavities.
\newblock {\em Nature}, 424:839--846, 2003.

\bibitem{Kippenberg:07}
Tobias~J. Kippenberg and Kerry~J. Vahala.
\newblock Cavity opto-mechanics.
\newblock {\em Opt. Express}, 15(25):17172--17205, 2007.

\bibitem{Cross08}
R.~Lifshitz and M.C. Cross.
\newblock Nonlinear dynamics of nanomechanical and micromechanical resonators.
\newblock {\em Reviews of nonlinear dynamics and complexity}, 1:1--50, 2008.

\bibitem{Quidant13}
Romain Quidant, Jan Gieseler, and Lukas. Novotny.
\newblock Thermal nonlinearities in a nanomechanical oscillator.
\newblock {\em Nature Physics}, 9:806--810, 2013.

\bibitem{Chaste12}
J.~Chaste, A.~Eichler, J.~Moser, G.~Cellabos, R.~Rurali, and A.~Bachtold.
\newblock A nanomechanical mass sensor with yoctogram resolution.
\newblock {\em Nature Nanotechnology}, 7:301--304, 2012.

\bibitem{Cleland02}
A.N. Clelan and M.L. Roukes.
\newblock Noise processes in nanomechanical resonators.
\newblock {\em J.~Appl. Phys.}, 92(5):2758--2769, 2002.

\bibitem{Buks06}
R.~Almog, S.~Zaitsev, O.~Shtempluck, and E.~Buks.
\newblock High intermodulation gain in a micromechanical duffing resonator.
\newblock {\em Appl. Phys. Lett.}, 88:213509, 2006.

\bibitem{Moser12}
Mark Dykman, editor.
\newblock {\em Fluctuating Nonlinear Oscillators: From Nanomechanics to Quantum
  Superconducting Circuits}, chapter~13.
\newblock Oxford Univrsity Press, 2012.

\bibitem{Wu87}
Ling-An Wu, Min Xiao, and H.~J. Kimble.
\newblock Squeezed states of light from an optical parametric oscillator.
\newblock {\em JOSA-B}, 4:1465--1476, 1987.

\bibitem{Marchesonl98}
Luca Gammaltoni, Peter Hangi, Peter Jung, and Fabio Marchesonl.
\newblock Stochastic resonance.
\newblock {\em Rev.Mod.Phys.}, 70:223, 1998.

\bibitem{dykman75a}
M.I. Dykman.
\newblock Theory of nonlinear nonequilibrium oscillators interacting with a
  medium.
\newblock {\em Zh.Eksp.Theor.Fiz}, 68:2082--2094, 1975.

\bibitem{dykman92}
M.I. Dykman and P.V.E. McClintok.
\newblock Power spectra of noise-driven nonlinear systems and stochastic
  resonance.
\newblock {\em Physica D}, 58:10--30, 1992.

\bibitem{dykman94}
M.I. Dykman, D.G. Luchinsky, and R.~Manella.
\newblock Supernarrow spectral peaks and high-frequency stochastic resonane in
  systems with coexisting periodic attractors.
\newblock {\em Physical Review E}, 49(2):1198--1215, 1994.

\bibitem{HoChan06}
C.~Stambaugh and H.B. Chan.
\newblock Supernarrow spectral peaks near a kinetic phase transition in a
  driven nonlinear micromechanical oscillator.
\newblock {\em Phys. Rev. Lett.}, 97:110602, 2006.

\bibitem{Stambaugh06}
C.~Stambaugh and H.B Chan.
\newblock Noise-activated switching in a driven nonlinear micromechanical
  oscillator.
\newblock {\em Phys. Rev.~B}, 73:172302, 2006.

\bibitem{Andre12}
Stephan Andre, Lingzhen Guo, Vittorio Peano, Michael Mathaler, and Gerd Schon.
\newblock Emission spectrum of the driven nonlinear oscillator.
\newblock {\em Phys. Rev.~A}, 85:053825, 2012.

\bibitem{Notomi05}
Masaya Notomi, Akihiko Shinya, Satoshi Mitsugi, Goh Kira, Eiichi Kuramochi, and
  Takasumi Tanabe.
\newblock Optical bistable switching action of {Si} high-$q$ photonic-crystal
  nanocavities.
\newblock {\em Opt. Express}, 13(7):2678--2687, 2005.

\bibitem{Cowan03}
A.~R. Cowan and J.~F. Young.
\newblock Optical bistability involving photonic crystal microcavities and fano
  line shapes.
\newblock {\em Phys. Rev.~E}, 68:046606, 2003.

\bibitem{JoannopoulosJo08-book}
John~D. Joannopoulos, Steven~G. Johnson, Joshua~N. Winn, and Robert~D. Meade.
\newblock {\em Photonic Crystals: Molding the Flow of Light}.
\newblock Princeton University Press, second edition, February 2008.

\bibitem{Walls80}
P.D. Drummond and D.F. Walls.
\newblock Quantum theory of optical bistability
  $\uppercase\expandafter{\romannumeral 1\relax}$. nonlinear polarisability
  model.
\newblock {\em Journal of Physics A: Math. and Gen.}, 13:725, 1980.

\bibitem{WangLi08}
Lei Wang and Baowen Li.
\newblock Thermal memory: A stroage of phononic information.
\newblock {\em Phys. Rev. Lett.}, 101:267203, 2008.

\bibitem{Notomi13}
Kengo Nozaki, Shinji Matsuo, Koji Takeda, Tomonari Sato, Eiichi Kuramochi, and
  Masaya Notomi.
\newblock Ingaas nano-photodetectors based on photonic crystal waveguide
  including ultracompact burried heterostructure.
\newblock {\em Optics Express}, 21:19022, 2013.

\bibitem{Otey10}
Clayton~R. Otey, Wah~Tung Lau, and Shanhui Fan.
\newblock Thermal rectification through vacuum.
\newblock {\em Phys. Rev. Lett.}, 104(15):154301, 2010.

\bibitem{Roberts11}
Nick~A. Roberts and D.G. Walker.
\newblock A review of thermal rectification observations and models in solid
  materials.
\newblock {\em Journal of Thermal Sciences}, 50:648--662, 2011.

\bibitem{lenert14}
Andrej Lenert, David~M. Bierman, Youngsuk Nam, Walker~R. Chan, Ivan Celanovic,
  Marin Soljacic, and Evelyn~N. Wang.
\newblock A nanophotonic solar thermophotovoltaic device.
\newblock {\em Nature Nanotechnology}, 9:126--130, 2014.

\bibitem{Noda07}
S.~Noda, M.~Fujita, and T.~Asano.
\newblock Spontaneous-emission control by photonic crystals and nanocavities.
\newblock {\em Nature Photonics}, 1:449--458, 2007.

\bibitem{Haus84:coupled}
H.~A. Haus.
\newblock {\em Waves and Fields in Optoelectronics}.
\newblock Prentice-Hall, Englewood Cliffs, NJ, 1984.
\newblock Ch. 7.

\bibitem{chinmay15}
Chinmay Khandekar, Adi Pick, Steven~G. Johnson, and Alejandro~W. Rodriguez.
\newblock Radiative heat transfer in nonlinear kerr media.
\newblock {\em Phys. Rev.~B}, 91:115406, 2015.

\bibitem{Rodriguez07:OE}
Alejandro Rodriguez, Marin Solja{\v{c}}i{\'{c}}, J.~D. Joannopulos, and
  Steven~G. Johnson.
\newblock $\chi^{(2)}$ and $\chi^{(3)}$ harmonic generation at a critical power
  in inhomogeneous doubly resonant cavities.
\newblock {\em Opt. Express}, 15(12):7303--7318, 2007.

\bibitem{Marin02}
Marin Soljacic, Mihai Ibanescu, Steven~J. Johnson, Yoel Fink, and J.D.
  Joannopoulos.
\newblock Optimal bistable switching in nonlinear photonic crystals.
\newblock {\em Phys.Rev.E.}, 66:055601, 2002.

\bibitem{Johnson01:cavities}
Steven~G. Johnson, Attila Mekis, Shanhui Fan, and J.~D. Joannopoulos.
\newblock Molding the flow of light.
\newblock {\em Computing Sci. Eng.}, 3(6):38--47, 2001.

\bibitem{Strogatz94}
Steven~H. Strogatz.
\newblock {\em Nonlinear Dynamics and Chaos}.
\newblock Westview Press, Boulder, CO, 1994.

\bibitem{dykman84}
M.I. Dykman and M.A. Krivoglaz.
\newblock Theory of nonlinear oscillators interacting with a medium.
\newblock {\em Soviet Physics Reviews}, 5, 1984.

\bibitem{Nawrodt12}
J.~Komma, C.~Schwarz, G.~Hoffman, D.~Heinert, and R.~Nawrodt.
\newblock Thermo-optic coefficient of silicon at $1550nm$ and cryogenic
  temperatures.
\newblock {\em Appl. Phys. Lett.}, 101:041905, 2012.

\bibitem{Zin14}
Z.~Lin, T.~Alcorn, M.~Loncar, S.G. Johnson, and A.W. Rodriguez.
\newblock High-efficiency degenerate four wave-mixing in triply resonant
  nanobeam cavities.
\newblock {\em Phys. Rev.~A}, 89:053839, 2014.

\end{thebibliography}

\end{document}